\documentclass[conference, a4paper]{IEEEtran}
\IEEEoverridecommandlockouts
\usepackage{cite}
\usepackage{url} 
\usepackage{amsmath,amssymb,amsfonts}
\usepackage{algorithmic}
\usepackage{graphicx}
\usepackage{textcomp}
\usepackage{xcolor}
\usepackage[symbol]{footmisc}
\def\BibTeX{{\rm B\kern-.05em{\sc i\kern-.025em b}\kern-.08em
    T\kern-.1667em\lower.7ex\hbox{E}\kern-.125emX}}
\begin{document}

\title{LMM-driven Semantic Image-Text Coding for Ultra Low-bitrate Learned Image Compression}

\author{\IEEEauthorblockN{Shimon Murai}
\IEEEauthorblockA{\textit{School of Fundamental} \\ \textit{Science and Engineering} \\
\textit{Waseda University}\\
Tokyo, Japan \\
octachoron@suou.waseda.jp}
\and
\IEEEauthorblockN{Heming Sun}
\IEEEauthorblockA{\textit{Faculty of Engineering} \\ \textit{Yokohama National University}\\
Kanagawa, Japan \\
sun-heming-vg@ynu.ac.jp}
\and
\IEEEauthorblockN{Jiro Katto}
\IEEEauthorblockA{\textit{School of Fundamental} \\ \textit{Science and Engineering} \\
\textit{Waseda University}\\
Tokyo, Japan \\
katto@waseda.jp}}

\maketitle

\begin{abstract}
Supported by powerful generative models, low-bitrate learned image compression (LIC) models utilizing perceptual metrics have become feasible. Some of the most advanced models achieve high compression rates and superior perceptual quality by using image captions as sub-information. This paper demonstrates that using a large multi-modal model (LMM), it is possible to generate captions and compress them within a single model. We also propose a novel semantic-perceptual-oriented fine-tuning method applicable to any LIC network, resulting in a 41.58\% improvement in LPIPS BD-rate compared to existing methods. Our implementation and pre-trained weights are available at https://github.com/tokkiwa/ImageTextCoding.
\end{abstract}

\begin{IEEEkeywords}
Learned Image Compression (LIC), Large Multi-modal Model (LMM), Latent Diffusion Model (LDM). 
\end{IEEEkeywords}

\section{Introduction}
Learned image compression (LIC) is one of the image compression methods where input images are transformed into latent variables and then encoded with an entropy model. It replaces the transformation used in conventional image compression with nonlinear transformations based on neural networks. The early works of LIC targeted pixel-level fidelity similar to traditional methods including JPEG and BPG. On the other hand, recent studies have introduced ultra low-bitrate compression models that focus on subjective fidelity. These models utilize generative models, such as Diffusion Models\cite{ldm} and GANs\cite{gan}, to restore details lost during compression.

 We note that recent models \cite{MISC,lei2023text,PerCo} utilize text information for ultra low-bitrate compression to leverage its high semantic compression ability. Existing models pass raw text to the decoder side \cite{MISC} or only apply traditional compression methods \cite{PerCo}, leaving room to improve text compression ability. 
 
 Inspired by recent studies of large language model-based data compression \cite{valmeekamLLMZipLosslessText2023, deletang2024language}, we propose a novel method to generate and compress text information within a single large multi-modal model. This enables us to improve existing LMM-driven image compression methods without any increase in model parameters. Experimental results show that we achieved more than a 65\% text compression ratio. 

We further focus on the fact that the existing method \cite{MISC} exploits fine-tuned existing LIC networks. Even though the final metric is perceptual loss, the LIC network in \cite{MISC} is trained with pixel-wise loss (Mean Squared Error). Instead, we train the LIC network with a novel perceptual and semantic-oriented loss functions, which resulted in a 42\% decrease in rate-distortion loss. (See section IV) 

To summarize our contribution, we (1) construct a new method to generate captions of an image and compress them within a single LMM model, (2) propose a semantic-perceptual loss to efficiently train low-bitrate LIC model, and (3) make our implementations and trained weights public to further research and reproducible experiment. 

\begin{figure}[t]
\centerline{\includegraphics[width = 9.3cm]{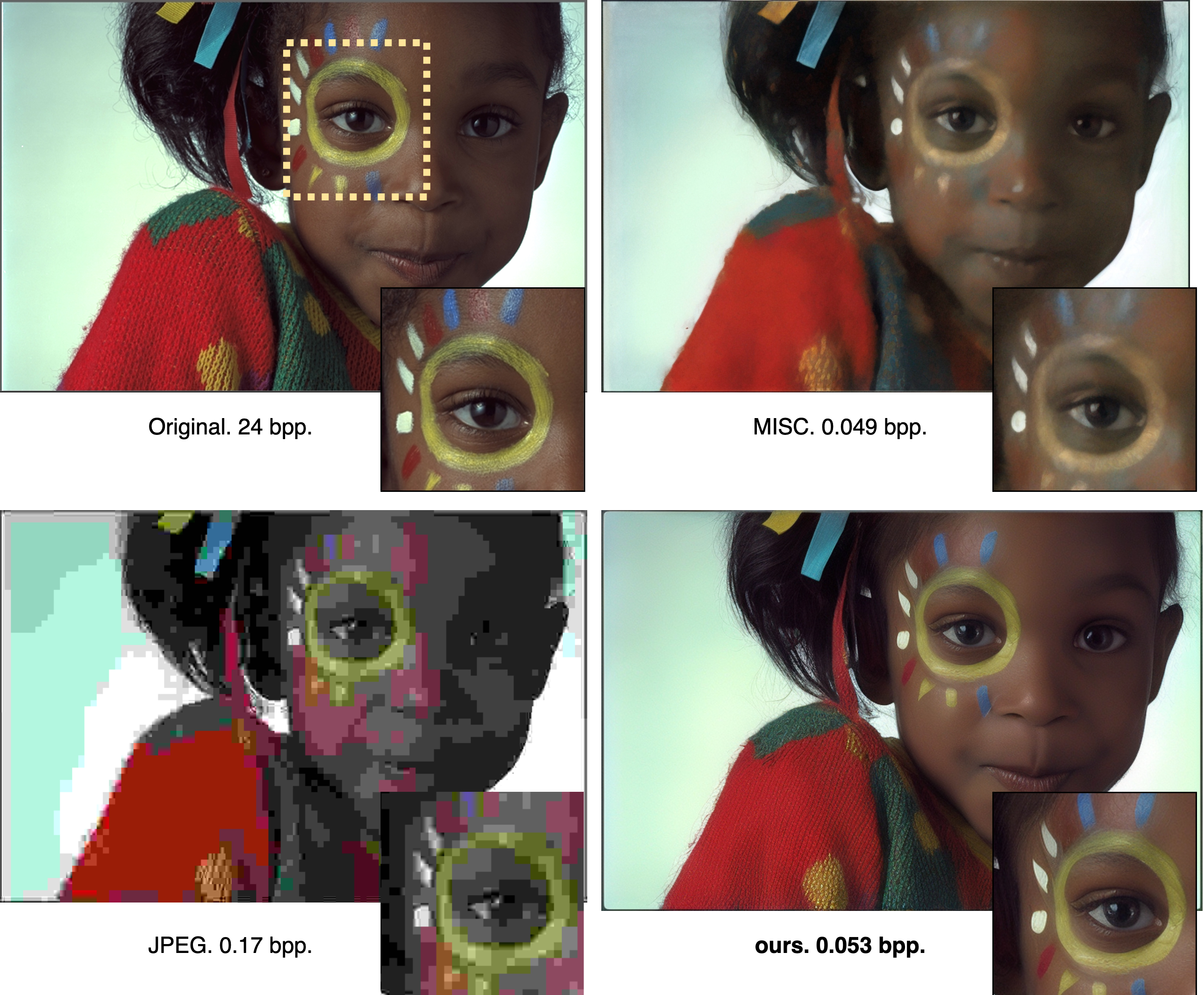}}
\caption{Compression results of Kodim15.png \cite{kodak} with a zoom-in. Our model achieves the ultra-low bitrate compression while avoiding the color distortion seen in MISC\cite{MISC}. }
\label{fig}
\end{figure}

\begin{figure*}[t]
 \begin{center}
  \includegraphics[height=7cm]{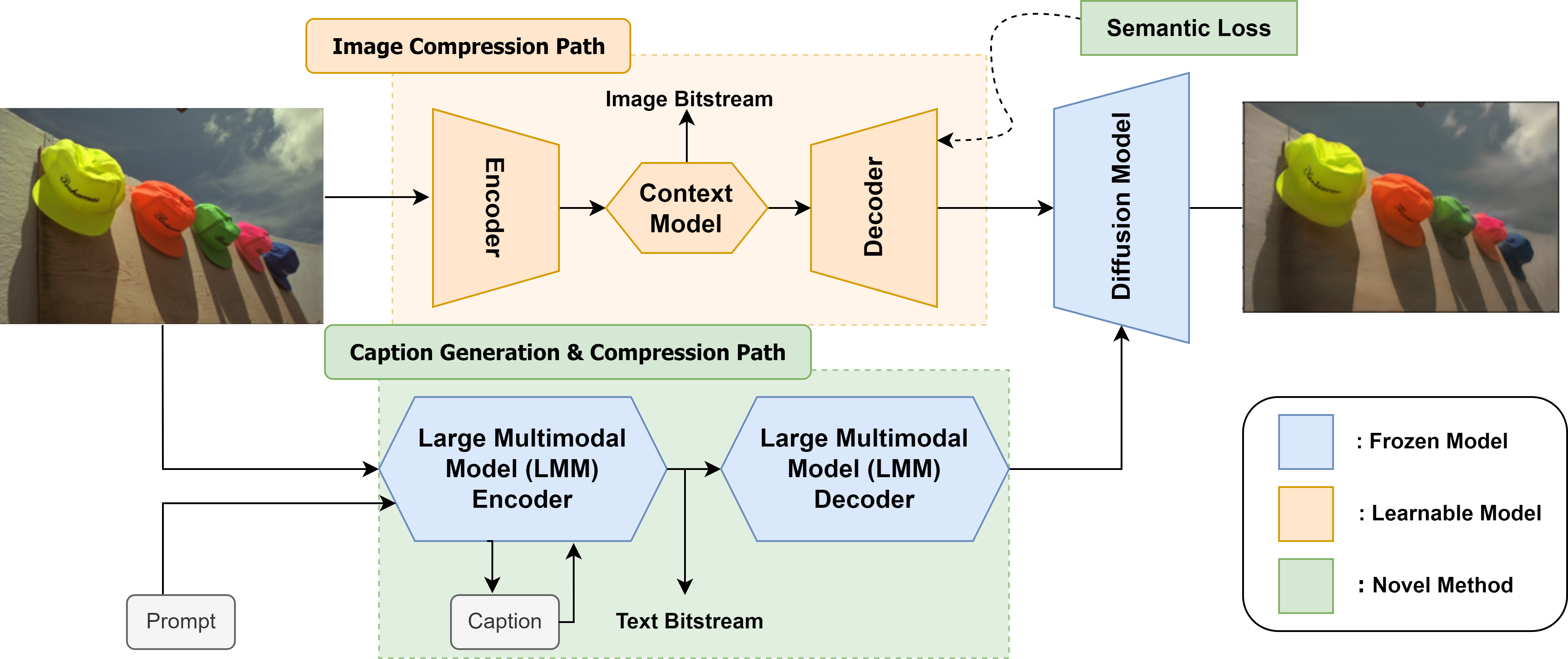}
  \caption{Our network architecture. The image is compressed to image bitstream with LIC model (above path), and at the same time, transported to LMM encoder (below path) to generate caption. The generated caption is then encoded to text bitstream. The two bitstreams are then decompressed and feeded to diffusion model. }
  \label{fig:arch}
 \end{center}
\end{figure*}

\section{Related Works}

\subsection{Generative Learned Image Compression}
Generative Learned Image Compression methods adopt generative models, such as Generative Adversarial Networks (GANs)\cite{gan} and Latent Diffusion Models (LDMs)\cite{ldm}, for the learned image compression task. These methods can generate perceptually good pictures at the expense of pixel-wise consistency\cite{survey}. In other words, generative image compression is superior to non-generative image compression in perceptual metrics such as FID\cite{fid} or LPIPS\cite{lpips}, while inferior to non-generative methods in pixel-wise metrics such as Mean Squared Error (MSE) or Peak Signal-to-Noise Ratio (PSNR). Restoring pixel-wise information usually requires more bit length, hence generative methods are suitable for ultra low-bitrate ($<$0.1 bpp) compression. The majority of existing ultra low-bitrate methods leverage GAN networks\cite{mentzerHighFidelityGenerativeImage2020,he2022po, iwai2021fidelity, agustsson2023multirealism} or LDM networks\cite{hoogeboomHighFidelityImageCompression2023,PerCo}.

\subsection{Text-Conditioned Learned Image Compression}
Text \& Sketch (or PICS)\cite{lei2023text} is one of the pioneering works to utilize text information for ultra low-bitrate image compression. Its encoder compresses caption text (in CLIP-embedding space) and edge information (sketch) of the input image, and its decoder reconstructs the image from the sketch with the help of a diffusion model conditioned by caption text. This enables ultra-low bitrate compression, yet the decompressed image only restores semantic features; the output image is quite different from the original image. To tackle this issue, MISC\cite{MISC} replaces the edge information with the full image. They compress the image with an existing neural image compression network which is fine-tuned to ultra-low bitrate. Then they decompress the image and apply four-step diffusion reconstruction with the caption text. The advantage of this model is that it contains an LMM model (GPT-4\cite{gpt4}), LIC model (Cheng\cite{cheng2020learned}), and diffusion model (DiffBIR\cite{diffbir}), which are replaceable with other kinds of architecture. 

\begin{figure}[t]
\centerline{\includegraphics[height=3.2cm]{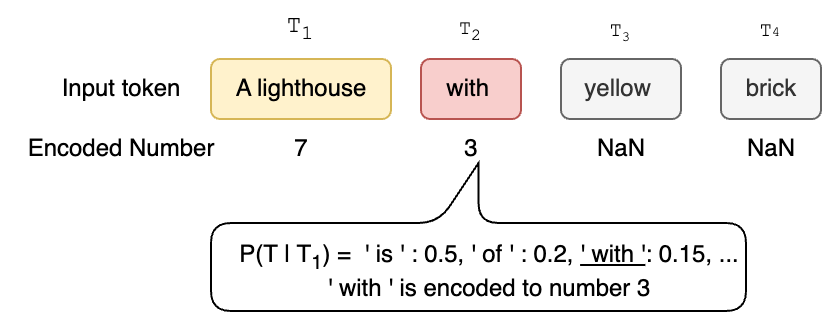}}
\caption{The visualization of LMM text compression.}
\label{lmmzip}
\end{figure}

\begin{figure*}[t]
\begin{minipage}[b]{0.45\linewidth}
\centerline{\includegraphics[height=6cm]{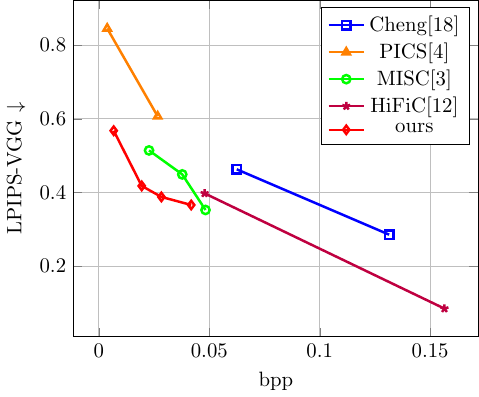}}
\caption{Relationship between bpp and LPIPS.}
\label{lpipsgraph}
\end{minipage}
\begin{minipage}[b]{0.45\linewidth}
\centerline{\includegraphics[height=6cm]{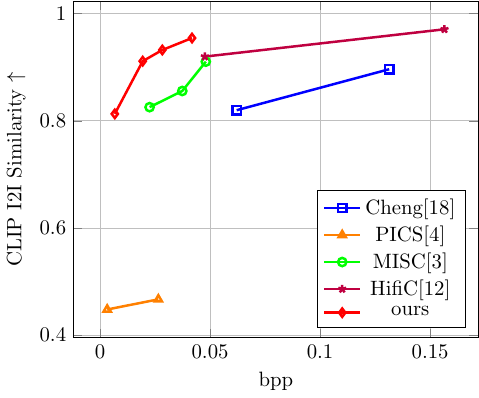}}
\caption{Relationship between bpp and CLIP Similarity.}
\label{clipsimgraph}
\end{minipage}
\end{figure*}

\subsection{Large Language Model and Large Multi-modal Model}
A large language model (LLM) is a network designed for conditional text generation. The LLM model is trained to predict the next token based on input tokens and given conditions. Recent researches \cite{llava} \cite{xu2024llava-uhd} realize the language models with image as an input. LLaVa \cite{llava}, a representative work of Large Multi-Modal Model(LMM),  projects the input image onto text embedding space via CLIP \cite{clip} encoder. 

Recent studies show large language models are suitable for data compression. The LLM models are able to predict the occurrence probabilities of the next token from given inputs, which are equivalent to lossless entropy coding. LLMZip\cite{valmeekamLLMZipLosslessText2023} computes the occurrence rank of the tokens and associates each token with the rank number. Another method \cite{deletang2024language} directly computes the probability of the next character and encodes it with an arithmetic encoder. 

\section{Proposed Method}
\subsection{Overview}

As shown in Figure \ref{fig:arch}, our framework consists of three components; (1) Image Compression Network, (2) Caption Generation \& Compression Network, and (3) Diffusion Network. In the encoding stage, the input image is passed into the image compressor to generate image bitstream. Meanwhile, the image is fed into the Caption-Compression Network, which outputs the compressed bitstream of the image caption. The decoder side decompresses the image and caption separately, to receive text caption and the reconstructed image. The decompressed image loses original information due to low-bitrate compression, so it is passed into the diffusion network, which we used the latent diffusion model, with text caption as a conditioning. 

Our network architecture is based on \cite{MISC}. The difference is that (1) we added the text compression path, (2) we added perceptual-semantic loss to the image coding path, and (3) we omitted the original three-stage diffusion process, as it does not contribute to the image quality despite that its heavy computational cost.

\subsection{LMM-driven text compression}

We adopt LLaVa \cite{llava} as our large multi-modal model. This model is constructed upon LLaMa \cite{llama}, enabling us to apply existing LLM-based compression methods \cite{valmeekamLLMZipLosslessText2023}. We first feed the model the input image and question template to receive the (tokenized) caption of the image. We then input the caption tokens again into the LMM model.

We show the visualization of this method in Fig. \ref{lmmzip}. The LMM model is trained so that, given the input token, it predicts the next token. For each token, we can receive probabilities of the next token, which are provided to the entropy coder (i.e. arithmetic coder) to encode the token. For ease of implementation, we convert the tokens into ranks and then encode the sequence of ranks with glib (adaptive Huffman coding). 

Rigorously, let $\lbrace T^{(1)}, \cdots T^{(N)}\rbrace$ denote the the tokens and $N$ denote the token length. For each token $T^{(i)}$, we first obtain conditional token occurrence probability $p(T^{(i)}| T^{(1)}, \cdots, T^{(i-1)})$ from the LMM model. Then, sort all possible $i$-th tokens $T^{(i)}_j$ by the probability and find where actual input $T^{(i)}$ is. The Rank $R^{(i)}$ is the index of $T^{(i)}$ in the sorted sequence. 

All tokens$\lbrace T^{(1)}, \cdots T^{(N)}\rbrace$ are transformed to Rank sequence $R^{(1)},\cdots,  R^{(N)}$. The sequence only contains small integers suitable for entropy coding. We adopt gzip compression to the sequence and obtain text bitstream.

\subsection{Semantic-Perceptual-Oriented Loss}
Our second contribution is to install new semantic and perceptual-oriented loss functions while training the LIC network. Previous work \cite{MISC} utilized only mean squared error (MSE) loss as a distortion metric, which is not suitable for the objective of high perceptual and semantic quality. 

Our loss function combines MSE and LPIPS-VGG\cite{lpips} as perceptual losses with additional CLIP-IQA\cite{clipiqa} loss and CLIP image-to-text score \cite{clip} as semantic losses. We abbreviate LIPIS-VGG as LPIPS and CLIP image-to-text to CLIPI2T for the following. CLIP-IQA loss is the cosine similarity between a given image and fixed prompt (e.g. 'good photo', 'real photo') both embedded with CLIP encoder. 
CLIP image-to-text score is also the cosine similarity between a given image and its caption, both embedded with CLIP encoder.
To combine all these metrics, we normalize them in the range $[0,1]$ by dividing them by their maximum value. For example, MSE loss takes the value between $[0,255]$, so we divide it by 255. Then define our loss function as follows: 
\begin{align}
    \text{Loss} &= \mathcal{R} + \lambda \mathcal{D}\\
    \mathcal{D} &= \kappa_0\mathcal{L}_\text{MSE} + \kappa_1\mathcal{L}_\text{LPIPS} + \kappa_2\mathcal{L}_\text{CLIPIQA} + \kappa_3\mathcal{L}_\text{CLIPI2T} 
\end{align}

where $\lambda$ specifies the target bitrate and each $\kappa_i$ balances the weights of each loss. $\mathcal{R}$ denotes the average bitrate, calculated through the entropy of the latent variable and estimated probability distribution.

\section{Experiments}
\subsection{Experimental Settings}
We froze the LMM network and Diffusion network and only focused on training the LIC network. We adopt pre-trained LLaVa-1.6-Mistral-7B \cite{llava} for the text generation-compression model and DiffBIR-v1 \cite{diffbir} for the generative reconstruction model. Detailed prompts and settings can be found in our GitHub repository. 

To generate the caption, we instructed the LMM network to give the description of the image within 50 words. In order to maintain compression ability, we did not select top-K search for generation, but greedy search. 
\subsection{Training}

The pre-trained (with MSE) Cheng\cite{cheng2020learned} model is fine-tuned with COCO \cite{lin2015microsoftcococommonobjects} 2014 dataset with approx. 80000 images and 50000 iterations. Each image is cropped to $256 \times 256$ and images smaller than cropping sizes are eliminated. Learning rate and batch-size were set to 1e-4 and 16 respectively. Each kappa was fixed to $( \kappa_0, \kappa_1, \kappa_2, \kappa_3 ) = (0.5, 0.2, 0.2, 0.1)$, and we trained four models with $\lambda = \lbrace 1, 2, 3, 4\rbrace$.

\subsection{Evaluation}

We tested our models with the CLIC2020-Professional \cite{clic2020} dataset. Note that this dataset is not labeled, hence not included in the training set of our multi-modal model. For the evaluation metric, we selected LPIPS \cite{lpips} and CLIP image-to-image similarity \cite{clip}. 

\paragraph{Image Compression Performance}
CLIP image-to-image similarity (CLIPI2I) is the cosine similarity of two images in CLIP-embedded space. As CLIP embedding space is shared with text,  we expect that a small distance in CLIP embedding space implies high semantic similarity. Hence we utilize this metric to evaluate semantic consistency. 
Experimental results are shown in Fig. \ref{lpipsgraph} and Fig. \ref{clipsimgraph}. Our method outperforms MISC\cite{MISC} and other methods in both CLIPI2I and LPIPS R-D curves. We calculated the Bj\o ndegaard-delta rate \cite{Bjntegaard2001CalculationOA} against MISC\cite{MISC}.  Our methods show 41.58 \% bitrate saving in LPIPS and 60.99 \% in Clip Image-to-Image Similarity. 

The qualitative result is shown in Fig. \ref{fig}. Our model eliminates the color distortion seen in MISC \cite{MISC} and JPEG while keeping an ultra-low bit-rate.

\paragraph{Text Compression Performance}
For the text compression, we computed the compression ratio with an LMM-generated caption of the CLIC2020 dataset. We instruct our LMM to give the description of the image in 50 words. Table \ref{textcompressionratio} shows that we achieve a 65.14\% compression ratio with LMM-driven rank encoding and gzip, which is an additional 21.86 \% gain to gzip-only compression.

\begin{table}[t]
\caption{Text Compression Ratio.}
\begin{center}
\begin{tabular}{|c|c|c|c|}
\hline
 & \textbf{\textit{None}}& \textbf{\textit{gzip}}& \textbf{\textit{Rank Encode + gzip}} \\
 \hline
 \textbf{\textit{Total bits}} & 1055856 & 598960 & 
$\boldsymbol{368080.0}$\\
\hline
 \textbf{\textit{Compression ratio}} & 0\% & 43.28\% & $\boldsymbol{65.14}$\% \\
 \hline
\end{tabular}
\label{tab1}
\end{center}
\label{textcompressionratio}
\end{table}

\section{Ablation Study}
\begin{figure}[t]
\centerline{\includegraphics[height=7.5cm]{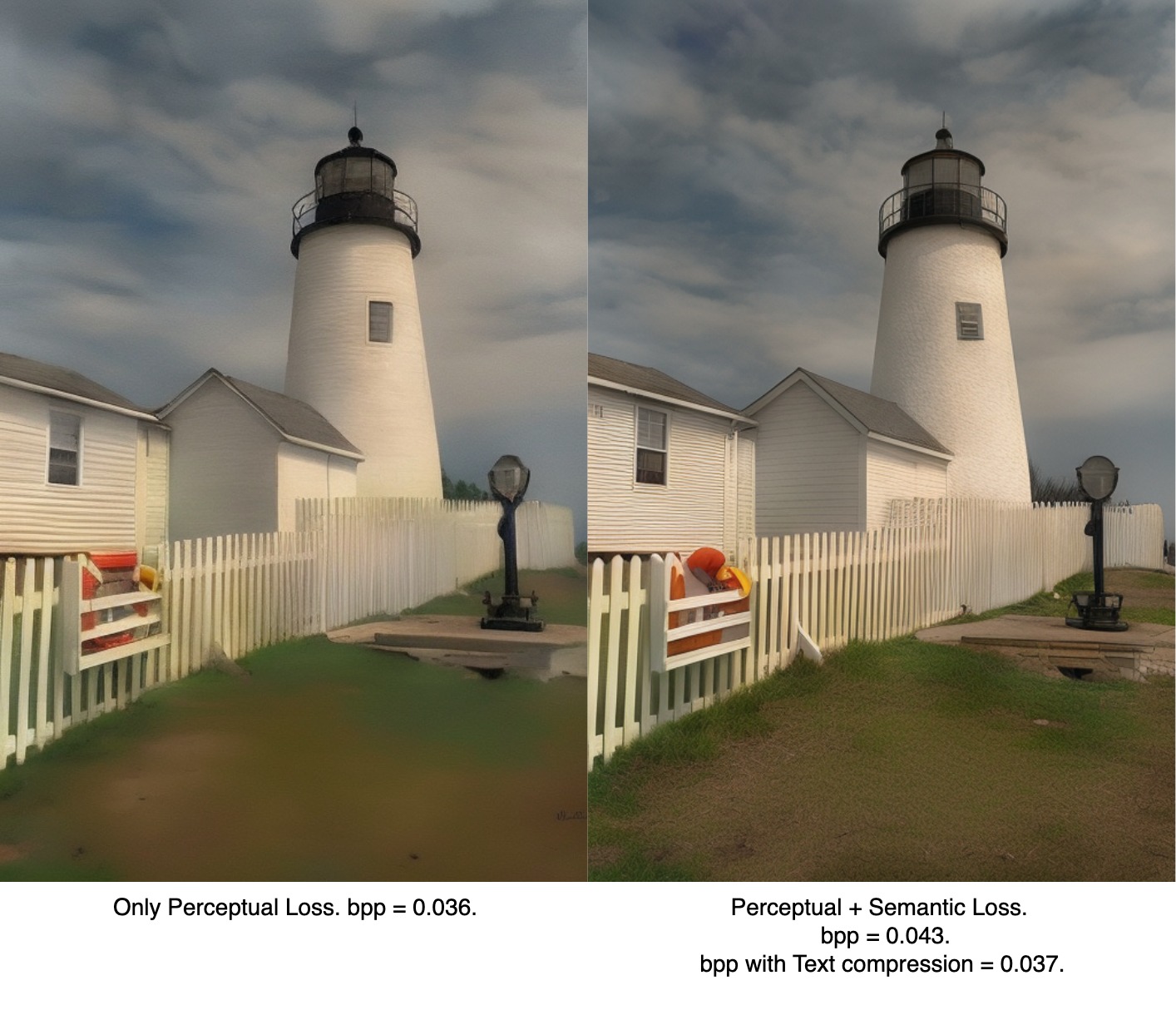}}
\caption{The visualization of kodim19.png\cite{kodak} encoded with the perceptual loss model (left) and our perceptual + semantic loss model (right).}
\label{perceptual_kodim15}
\end{figure}

In order to evaluate the effectiveness of our semantic-perceptual-loss, we trained another variant with only perceptual-loss. We set $\kappa_0 = \kappa_1 = 0.5$ and $\kappa_2 = \kappa_3 = 0$ to disable semantic loss. The result in Table \ref{bdrate} shows that the difference in LPIPS-bpp and CLIPI2I-bpp is not significant. On the other hands, only perceptual-loss model suffers from the noise seen  some images. (Fig. \ref{perceptual_kodim15}) This phenomenon is caused because the intermediate image is overly optimized to LPIPS loss. 

In the case of training at ultra-low bitrates, the use of Neural Network-based losses such as LPIPS\cite{lpips} may result in over-fitting to that metric, producing distorted images. As a related example, PO-ELIC\cite{he2022po} aims to improve subjective quality by combining multiple loss metrics. Adding semantic loss is expected to prevent over-fitting to a single metric and contribute to subjective quality.

Furthermore, we show the comparison of all methods, including perceptual loss, perceptual and semantic loss, and text compression in Table \ref{bdrate}. Our model overwhelm the existing method \cite{MISC} in both LPIPS and CLIP image-to-image Similarity with more than 40 \% saving of bitrate. 

\begin{table}[t]
\caption{BD-rate \cite{Bjntegaard2001CalculationOA} for each method. MISC\cite{MISC} as Anchor.}
\begin{center}
\begin{tabular}{|c|c|c|}
\hline
  & \textbf{\textit{LPIPS}}& \textbf{\textit{CLIP I2I}}\\
 \hline
 \textbf{\textit{Perceptual}} & -39.76\%  & -57.38\%  \\
\hline
 \textbf{\textit{Perceptual + Semantic}} & -38.07\% & -57.47\%  \\
 \hline
  \textbf{\textit{Perceptual + Semantic}} & $\boldsymbol{-41.58}\%$ & $\boldsymbol{-60.99}\%$  \\
  \textbf{\textit{ + Text Compression}} &  &   \\
 \hline
\end{tabular}
\end{center}
\label{bdrate}
\end{table}

\section{Conclusion}
In this paper, we have presented a novel approach to ultra low-bitrate image compression by integrating caption generation and compression within a single large multi-modal model (LMM). Our method leverages the semantic richness of text descriptions to enhance the perceptual quality of compressed images. By incorporating a semantic-perceptual-oriented loss function, we further improve the compression performance, achieving significant gains in rate-distortion metrics compared to existing methods.

Future work may explore the extension of this framework to other modalities and applications, as well as the refinement of the compression model for even greater compression efficiency. 

\section*{Acknowledgment}
This work is supported in part by SCAT, in part by JSPS KAKENHI Grant Number JP23K16861, and in part by Telecommunications Advancement Foundation. 

\bibliography{lmmlic.bib} 
\bibliographystyle{ieeetr}

\end{document}